\documentclass[11pt]{article}

\usepackage[utf8]{inputenc}
\usepackage[T1]{fontenc}

\usepackage[margin=1in]{geometry}
\usepackage{url}
\usepackage{booktabs}
\usepackage{amsfonts}
\usepackage{nicefrac}
\usepackage{microtype}
\usepackage{xcolor}
\usepackage{graphicx}
\usepackage{amsmath}
\usepackage{listings}
\usepackage[numbers,compress]{natbib}
\usepackage{caption}

\begin{document}

\title{NL2SQL with Iterative Self-Refinement}

\author{
  Anjali Vemula \\
  Department of CSE \\
  University of Minnesota - Twin Cities \\
  Minneapolis, MN 55455 \\
  \texttt{vemul034@umn.edu}
  \and
  Akshat Dasula \\
  Department of CSE \\
  University of Minnesota - Twin Cities \\
  Minneapolis, MN 55455 \\
  \texttt{dasul001@umn.edu}
}
\date{}
\maketitle

\begin{abstract}

  Bridging the gap between human intent and machine logic remains a critical challenge in data access for non-technical users, yet Large Language Models (LLMs) remain unreliable: they hallucinate schema elements, generate syntactically invalid queries, or fail to capture user intent. These limitations are especially critical in domains such as insurance, where KPIs, regulatory constraints, and legacy relational systems demand high precision. We introduce a multi-agent, persona-driven NL2SQL framework that combines (1) domain relevance detection, (2) schema-grounded knowledge construction, (3) KPI-aware SQL synthesis, (4) syntax-level self-repair using sqlite execution feedback, and (5) semantic consistency checking via SQL-to-NL reverse engineering. Our system significantly reduces query hallucination and improves semantic fidelity compared to direct LLM prompting. Evaluated on 60 queries (50 from a HuggingFace text-to-SQL dataset and 10 custom domain queries), the system achieves 90\% correctness using Mistral across seven coordinated LLM calls showing that this approach can substantially increase the reliability of NL2SQL pipelines in high-stakes environments.
\end{abstract}

\section{Introduction}
The insurance industry operates on complex, data-intensive processes that rely heavily on structured information about policies, claims, customers, agents, and submissions. These datasets carry critical business functions, including pricing, underwriting, fraud detection, portfolio analysis, and KPI-driven performance monitoring. Although insurers store decades of transactional data in relational databases, this information is often locked behind SQL interfaces accessible only to technically trained analysts. As a result, underwriters, financial planners, and operations teams depend on technical intermediaries to obtain even basic metrics, delaying decisions in a domain where timing and accuracy are crucial.

Despite increasing adoption of AI tools, querying these legacy systems remains challenging \cite{survey}. Their schemas are domain-specific, include numerous interlinked tables, and require precise understanding of insurance KPIs such as loss ratios or claim frequencies. Modern LLMs like ChatGPT can translate natural language into SQL, but they frequently hallucinate table or column names, produce syntactically invalid queries, or generate outputs misaligned with business intent. Such unreliability makes them unsuitable for high-stakes industries like insurance, where incorrect queries can lead to flawed reports, mispriced risk, or regulatory compliance issues.

To address these challenges, we develop a multi-agent, persona-driven NL2SQL system tailored for insurance databases. Our approach integrates (1) domain-relevance validation to ensure user queries align with insurance concepts (2) schema-grounded knowledge generation \cite{align}, including KPI decomposition (3) iterative SQL correction using sqlite-based syntax checks and (4) semantic validation via SQL-to-NL reverse engineering paired with LLM-as-a-judge comparison. By combining domain expertise, guided decomposition, and error-correcting feedback loops, our system produces SQL that is both syntactically correct and semantically faithful to user intent. Through this structured methodology, we demonstrate that reliable natural-language querying of legacy insurance databases is achievable, overcoming the limitations of direct LLM prompting and enabling decision-makers to access critical insights without technical barriers.

\section{Related Work}
Research in text-to-SQL has evolved through specialized architectures, prompting strategies, and evaluation frameworks aimed at improving the reliability of NL2SQL systems. Early work focused on neural architectures that explicitly model schema-linking and relational structures. For example, RAT-SQL \cite{ratsql} introduced a relation-aware encoder that aligns natural-language queries with database schemas through semantic linking, achieving strong performance on cross-domain benchmarks. This class of models demonstrates the importance of grounding language understanding in structural representations of the database.

As large language models (LLMs) became more prominent, prompting strategies emerged as a practical alternative to heavy model training. Techniques such as Divide-and-Prompt and Chain-of-Thought (CoT) prompting \cite{cottxt2sql} have shown that breaking NL queries into smaller reasoning steps can improve SQL generation accuracy. These methods emphasize guiding the model toward intermediate reasoning, reducing some forms of hallucination and strengthening schema awareness.

It is necessary to use  evaluation frameworks such as LLM-as-a-Judge or SQL2NL \cite{revengg} comparison to provide more reliable assessments of semantic equivalence than simple string matching. These methods compare the intent of the NL query with the intent expressed by the generated SQL, offering a closer proxy for actual correctness.

However, even with advanced prompting, generating accurate SQL remains difficult. The base paper influencing our work is SQL-of-Thought \cite{sqlthought} which proposed a multi-agent, error-correcting pipeline where LLMs iteratively refine SQL queries based on execution feedback but this approach only relied on syntactic correctness of the generated query.

Unlike prior NL2SQL systems that rely solely on schema-linking architectures, prompting techniques, or single-pass SQL generation, our approach introduces a persona-driven, KPI-aware, multi-stage correction pipeline specifically tailored for the insurance domain. We extend beyond traditional parsing by incorporating domain relevance filtering, explicit knowledge generation (including tables, columns, and KPI formulas), and a guided refinement loop that uses both syntactic validation through SQLite execution and semantic validation via SQL-to-NL reverse engineering.

\section{Insurance Database}

\begin{figure}
    \centering
    \includegraphics[width=1\linewidth]{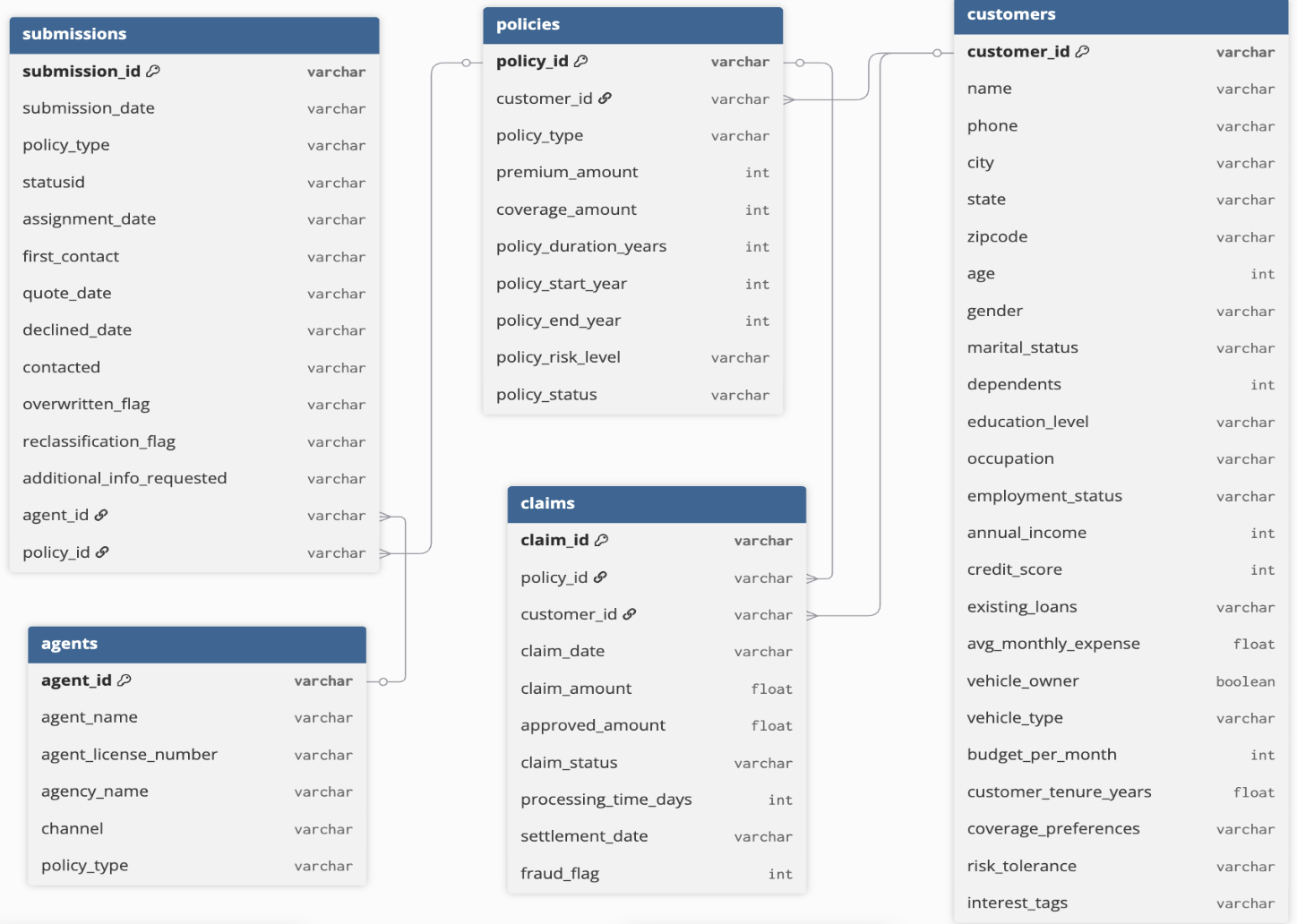}
    \caption{Insurance Database Schema}
    \label{fig:schema}
\end{figure}

To evaluate and execute queries generated by our persona-driven NL2SQL framework, we created a synthetic database as shown in \ref{fig:schema}, specifically tailored to the insurance sector. This database included complex, domain-specific relational structures that mimics real-world insurance. We embedded the below information in the prompts using TOON \cite{toon_format}

\subsection{Schema Overview}
The database models a standard insurance lifecycle, comprising five core relational entities that track the customer journey from lead acquisition to policy management and claims processing.

\begin{itemize}
  \item \textbf{\texttt{customers}}: Serves as the central dimension table containing detailed demographic and financial profiles, including attributes such as \textit{\texttt{annual\_income}, \texttt{credit\_score}, \texttt{risk\_tolerance},} and \textit{\texttt{marital\_status}}. This table enables fine-grained segmentation for risk analysis.

  \item \textbf{\texttt{policies}}: Records the contractual details of issued insurance products. Key attributes include \textit{\texttt{premium\_amount}, \texttt{coverage\_amount}, \texttt{policy\_risk\_level},} and temporal fields like \textit{\texttt{policy\_start\_year}} and \textit{\texttt{policy\_end\_year}}, allowing for longitudinal portfolio analysis.

  \item \textbf{\texttt{claims}}: A transactional table tracking liabilities. It logs \textit{\texttt{claim\_amount}, \texttt{approved\_amount}, \texttt{fraud\_flag}}, and \textit{\texttt{processing\_time\_days}}, which are critical for calculating loss ratios and fraud detection metrics.

  \item \textbf{\texttt{agents}}: Stores metadata on the salesforce, including \textit{\texttt{agent\_license\_number}, \texttt{agency\_name}}, and sales \textit{\texttt{channel}}. This table is essential for performance attribution and channel efficacy reporting.

  \item \textbf{\texttt{submissions}}: Represents the early-stage funnel of insurance quotes and leads. It tracks the status of applications (e.g., \textit{\texttt{statusid}} for \textit{``Bound''} vs.\ \textit{``Quote Only''}) and links prospective policies to specific agents.
\end{itemize}

\subsection{Domain Knowledge Representation}

To strictly bound the model's reasoning and mitigate the hallucination of non-existent columns or incorrect metric derivations, we augmented the input context with two structured metadata dictionaries.

\textbf{KPI Dictionary:} To ensure mathematical precision, we embedded a predefined dictionary of Key Performance Indicators (KPIs) that maps industry terms to their exact formulae. Instead of relying on the LLM’s knowledge to define metrics such as ``loss ratio'' or ``conversion rate,'' the system retrieves explicit formulas (e.g., $\text{Quote Conversion} = \frac{\text{Count(Bound)}}{\text{Count(Quoted)}}$). This ensures that aggregate metrics align with business logic rather than probabilistic approximations. Few example KPIs are shown below:

\begin{lstlisting}[label={lst:kpi_dictionary}]
"customer_lifetime_value": {
  "meaning": "Total premium revenue expected from a customer over their entire relationship with the insurer.",
  "math_formula": "SUM(premium_amount) WHERE policy.customer_id = customer.customer_id"
},
"premium_to_income_ratio": {
  "meaning": "Percentage of customer income spent on insurance premiums. Indicates affordability.",
  "math_formula": "(premium_amount / annual_income) * 100"
}
\end{lstlisting}

\textbf{Enriched Schema Dictionary:} To prevent schema misalignment, we provide a comprehensive schema dictionary that details every relation. Beyond standard table and column names, this dictionary includes:
  \begin{itemize}
    \item \textit{Semantic descriptions} that explain the business meaning of each attribute to resolve ambiguities (e.g., distinguishing between \textit{policy\_start\_year} and \textit{assignment\_date}).
    \item \textit{Data types} specifying explicit type constraints (e.g., string, int, boolean) to ensure correct operator usage.
    \item \textit{Column statistics} (e.g., value ranges and categorical cardinality), enabling the model to understand valid distributions. This is crucial for generating accurate WHERE clauses and avoiding queries over non-existent values.
  \end{itemize}

An example of the relation is shown below:


\begin{lstlisting}[label={lst:schema_dictionary}]
{
  "table_name": "agents",
  "columns": {
    "agent_id": "Unique identifier for each agent, format 'A' followed by a number (e.g., A1000), categorical",
    "agent_name": "Full name of the agent (string)",
    "agent_license_number": "Unique license number for the agent, format 'LIC-' followed by 8-character alphanumeric code (string)",
    "agency_name": "Name of the agency the agent belongs to (string)",
    "channel": "The sales channel the agent operates in, categorical, values: ['Telemarketing', 'Digital', 'Branch']",
    "policy_type": "Line of business the agent handles, categorical, values: ['auto', 'home', 'health', 'life']"
  }
}
\end{lstlisting}

\section{Methodology}

\begin{figure}
    \centering
    \includegraphics[width=0.65\linewidth]{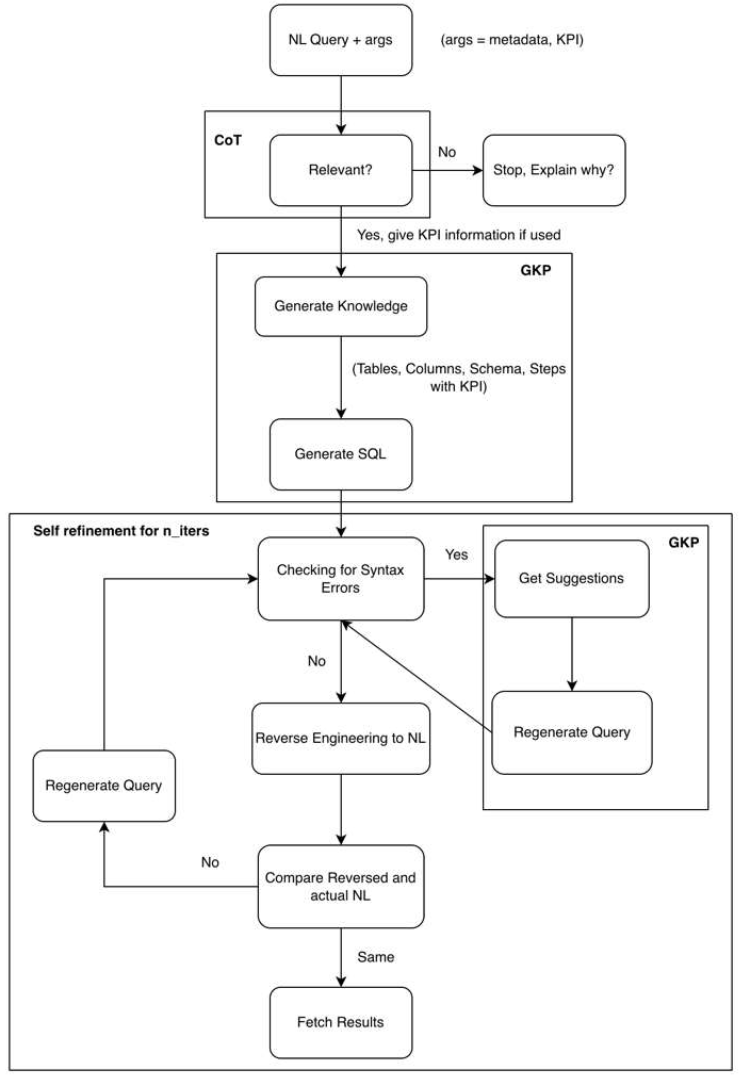}
    \caption{Flow of Execution}
    \label{fig:flow}
\end{figure}

We define the Text-to-SQL task as the generation of an executable SQL query $s$ given a natural language query $q$ and a database schema $S$. Unlike standard one-shot generation approaches, we model this process as a multi-stage iterative system, as shown in \ref{fig:flow} comprising an initial generation phase followed by a self-refinement loop driven by dual feedback mechanisms.

\subsection{Query Generation}
The process begins with a \textit{Generation Step}, denoted as $f_{\text{gen}}$, which synthesizes an initial SQL query candidate $s_0$. This function utilizes an intermediate knowledge representation $K(q, S)$, comprising relevant tables, columns, and KPI formulas derived from the schema $S$ rather than mapping directly from $q$.

Formally, the initial state is defined as:
\begin{equation}
    s_{0} = f_{\text{gen}}(q, K(q, S), S)
\end{equation}
where:
\begin{itemize}
    \item $q$ represents the Natural Language (NL) query.
    \item $S$ represents the database schema and metadata.
    \item $K(q, S)$ denotes the generated guidance (schema linking, KPI decomposition, and reasoning steps).
\end{itemize}

Before generating the query, we first check the relevance of each user query to the insurance domain using Chain-of-Thought prompting \cite{cot}. For example, when asked about a loss ratio, the system produces the following explanation:

\textbf{Query:} \textit{``What is the loss ratio in 2023 and 2024?''}

\textbf{Relevance Explanation:} \textit{``The query asks for the loss ratio, which directly refers to a key performance indicator (KPI) in the insurance domain.''}

In contrast, when presented with a query that is unrelated to the domain, the system identifies the mismatch:

\textbf{Query:} \textit{``What is the cost of this swimming pool?''}

\textbf{Relevance Explanation:} \textit{``The query asks for the cost of a swimming pool, which is not related to the insurance domain.''}

If the query not relevant, the generation is stopped here. Otherwise, the next step is to generate knowledge \cite{gkp} relevant to the query by embedding KPI and schema information in the prompt. An example of the generated knowledge is given below:

\begin{lstlisting}[label={lst:knowledge_op}]
{
  "Tables": ["submissions", "agents"],
  "Columns": {
    "submissions": ["statusid", "agent_id"],
    "agents": ["agent_id", "channel"]
  },
  "Steps": [
    "Join the 'submissions' table with the 'agents' table using the 'agent_id' column.",
    "Filter the submissions where 'statusid' is either 'BND' (Bound) or 'QTE' (Quote Only).",
    "Group the results by the 'channel' column from the 'agents' table.",
    "Count the total number of quotes (statusid = 'QTE') and the number of quotes that are bound (statusid = 'BND') for each channel.",
    "Calculate the quote conversion rate using the KPI formula: (COUNT(quotes_bound) / COUNT(total_quotes)) * 100.",
    "Select the 'channel' and the calculated quote conversion rate, ensuring the result is in float numbers."
  ]
}
\end{lstlisting}

This knowledge is then used to generate the relevant SQL query which is then passed to the Query Execution step.

\subsection{Query Execution and Self-Refinement}

To ensure both syntactic correctness and semantic alignment, the system enters a \textit{Refinement Step} \cite{refine}, denoted as $f_{\text{ref}}$. This process iteratively updates the SQL query $s_t$ at step $t$ by incorporating feedback from two distinct error reflection functions: $E_{\text{syn}}$ (syntax error reflection) and $E_{\text{log}}$ (logical error reflection).

The state transition for the query refinement is defined recursively as:
\begin{equation}
    s_{t+1} = f_{\text{ref}}(q, s_{t}, E_{\text{syn}}(s_{t}), E_{\text{log}}(s_{t})) \quad \text{for } t = 0, 1, \dots, N
\end{equation}
where:
\begin{itemize}
    \item $s_t$ is the generated SQL query at iteration $t$.
    \item $E_{\text{syn}}(s_t)$ represents the syntax error feedback obtained from executing $s_t$ on the target database.
    \item $E_{\text{log}}(s_t)$ represents the logical error reflection derived from reverse-engineering $s_t$ back into natural language and comparing it against $q$.
    \item $N$ is the maximum number of allowed refinement iterations.
\end{itemize}

This formulation ensures that $s_{t+1}$ is strictly conditioned on the specific failure modes of $s_t$, allowing the model to converge toward a solution that is both executable and semantically faithful to the user's intent.

The generated SQL query is executed on the sqlite3 database, if the query is executed, it shows that the query is syntactically correct and it is then moved to logical error analysis phase. If there are syntactic errors, the self-refinement loop begins by passing these errors and the SQL query as another LLM call for suggestions to correct the query. These suggestions are then used to generate the corrected query. An example of syntax error refinement is given below:

\begin{lstlisting}[label={lst:syntax_correction}]
NL Query: What is the quote to bind ratio per agent channel? Use float numbers.

SQL Query:
SELECT a.channel,
       (COUNT(CASE WHEN s.statusid = 'BND' THEN 1 END) /
        COUNT(CASE WHEN s.statusid = 'QTE' THEN 1 END) * 100)::float
       AS quote_conversion_rate
FROM submissions s
JOIN agents a ON s.agent_id = a.agent_id
WHERE s.statusid IN ('BND', 'QTE')
GROUP BY a.channel

Passing Syntax Error Trial: 1
------------------------------------------------------------
Syntax Error: unrecognized token: ":"
Reflecting on Syntax Error....

Syntax OP:
The error you're encountering is due to the misuse of the ::float cast operator. In SQL, the ::float syntax is specific to certain dialects like PostgreSQL. For more general SQL compatibility, you should use the CAST function.

### Explanation of Changes:

1. Casting to Float:
   - Changed ::float to CAST(... AS FLOAT) for better compatibility across different SQL dialects.
   - Multiplied the count of bound quotes by 1.0 to ensure the division results in a float rather than an integer.

2. Ensuring Safe Division:
   - To avoid division by zero, you might want to add a check to ensure there are quotes (statusid = 'QTE'). However, since the filtering already ensures that only records with statusid either 'BND' or 'QTE' are considered, this might not be strictly necessary.
\end{lstlisting}

In the logical error analysis phase, the system evaluates the semantic correctness of the generated SQL query using a reverse engineering strategy. Specifically, the SQL query is translated back into a structured natural language description that captures its intent, constraints, and aggregation logic. This generated explanation is then systematically compared against the user’s original natural language query using an LLM-as-a-Judge \cite{judge} mechanism, where a large language model scores semantic alignment between the two representations.

If the reverse-engineered description is judged to be semantically consistent with the user’s intent, the SQL query is executed against the target database and the results are returned. If a semantic mismatch is detected, the system automatically triggers a correction routine that produces a revised SQL query. This corrected query is fed back into a self-refinement loop, enabling iterative improvement through repeated generation, validation, and repair. This feedback-driven process helps prevent silent logical failures, reduces hallucinated joins or filters, and ensures that the final query reflects both the syntactic structure and the semantic intent of the user’s request.

\begin{lstlisting}[label={lst:reverse_engineering}]
The SQL query calculates the quote-to-bind ratio for different agent channels in the insurance domain. It does this by first joining the submissions table with the agents table on the agent_id field. It then filters the submissions to include only those with a status of 'BND' (bound) or 'QTE' (quoted). The query counts the number of bound submissions and the number of quoted submissions for each agent channel. It calculates the ratio of bound submissions to quoted submissions, converts this ratio to a percentage, and groups the results by the agent channel. The final output is the quote-to-bind ratio for each channel, indicating the effectiveness of different agent channels in converting quotes to bound policies.
{
  "correct": "No",
  "corrected_sql_query": "SELECT agents.channel,
    (CAST(COUNT(CASE WHEN submissions.statusid = 'BND' THEN 1 END) AS FLOAT) /
     NULLIF(CAST(COUNT(CASE WHEN submissions.statusid IN ('BND', 'QTE') THEN 1 END) AS FLOAT), 0))
     * 100 AS quote_to_bind_ratio
   FROM submissions
   JOIN agents ON submissions.agent_id = agents.agent_id
   WHERE submissions.statusid IN ('BND', 'QTE')
   GROUP BY agents.channel"
}
Logical Error....Retrying
Passing Syntax Error Trial: 2
------------------------------------------------------------
No Syntax Error....
------------------------------------------------------------
Passing Logical Error Trial: 2
------------------------------------------------------------
{
  "correct": "Yes",
  "corrected_sql_query": "None"
}
No Logical Error....
\end{lstlisting}

\section{Results and Discussion}

\begin{figure}[h]
    \centering
    \begin{minipage}{0.48\linewidth}
        \centering
        \includegraphics[width=\linewidth]{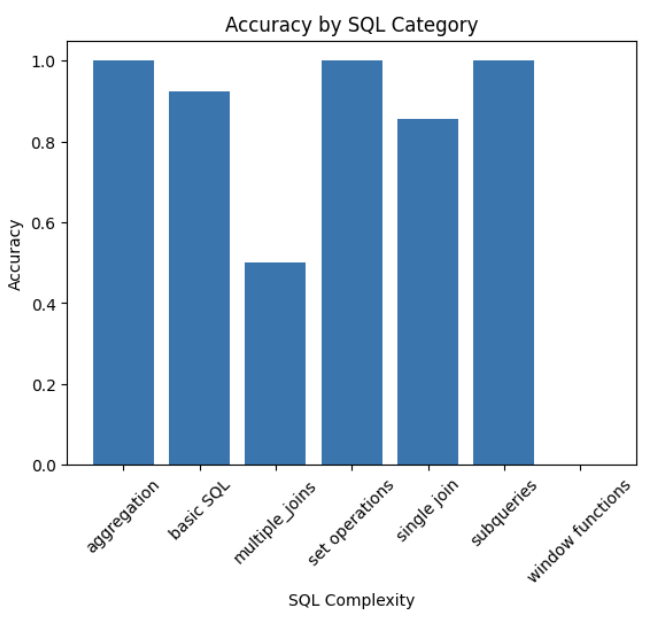}
        \caption{Accuracy breakdown by SQL Complexity}
        \label{fig:results}
    \end{minipage}\hfill
    \begin{minipage}{0.48\linewidth}
        \centering
        \includegraphics[width=\linewidth]{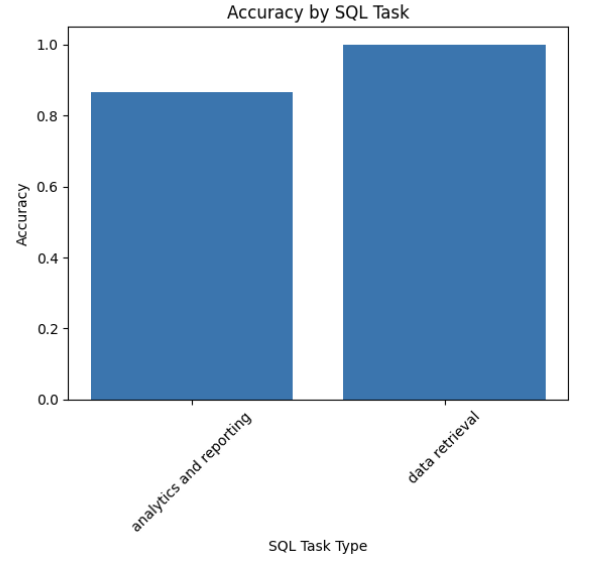}
        \caption{Accuracy breakdown by SQL Task Type}
        \label{fig:results2}
    \end{minipage}
\end{figure}

Our experimental evaluation on a curated set of 60 domain-specific queries comprising 50 instances from \texttt{gretelai} synthetic dataset\cite{data} and 10 custom KPI based complex queries, achieving an aggregate accuracy of approximately 90\%. Performance stratification by task type reveals that the system attains near-perfect reliability (100\%) in ``Data Retrieval'' tasks, which primarily involve direct selection and filtering. In contrast, the more abstract ``Analytics and Reporting'' scenarios, which require multi-step aggregation and KPI derivation, maintained a robust accuracy of approximately 85\%. Furthermore, the model exhibits resilience across varying levels of SQL complexity as shown in \ref{fig:results}, handling basic queries with high precision while maintaining substantial accuracy in generating advanced structures involving subqueries and window functions.


This system used many personas in its verification process such as ``Insurance Domain Specialist'' and ``SQL Data Analyst'' personas to iteratively tune queries that initially fail execution or logic checks. This multi-agent collaboration ensures that the final output is not only syntactically valid but semantically aligned with complex insurance logic.

Regarding operational constraints, the framework utilizes the \texttt{mistral-large-2411} model \cite{mistral} to power its reasoning agents. The pipeline consisting of Chain-of-Thought relevance reasoning, knowledge generation, syntax validation, and logical refinement, necessitates an average of \textbf{7 LLM calls} per query execution. This results in an average end-to-end latency of approximately 20 seconds. While this latency is higher than direct text-to-SQL inference, it produces accurate and reliable queries without hallucinating relations, attribute names, etc.

This work can be easily reproduced, please use this Github repo of NL2SQL \footnote{GitHub: \url{https://github.com/AkshatMittu/NL2SQL}}

\section{Conclusion and Future Work}
In this work, we presented a persona-driven, KPI-aware framework for Natural Language to SQL (NL2SQL) generation, specifically designed to address the accessibility gap in the insurance domain. By formalizing the generation process as a recursive function with dual-feedback loops, we successfully mitigated common LLM failure modes, including schema hallucination and syntactic invalidity. Our approach, which couples "Generate Knowledge Prompting" (GKP) with a rigorous "Reverse Engineering" validation step through self-refinement, allows non-technical users to reliably query legacy databases without knowledge of the underlying SQL structure. Experimental results demonstrate that this iterative refinement strategy achieves approximately 90\% accuracy, proving that agentic workflows can effectively bridge the divide between human intent and machine logic in high-stakes enterprise environments.

While our current implementation establishes a robust baseline, we envision several  enhancements to improve the framework's utility. Future iterations of this work will transition from static prompting to dynamic prompt modulation, automatically adjusting Chain-of-Thought depth and employing few-shot dynamic example selection based on query complexity. Concurrently, we intend to assess model agnosticism by benchmarking across many LLMs ranging from lightweight open-source models to proprietary frontiers to optimize the trade-off between latency, inference cost and reasoning accuracy. Finally, we plan to extend the pipeline’s capabilities beyond mere retrieval by integrating a post-processing agent that summarizes SQL results into actionable natural language insights, thereby transforming the system into a comprehensive decision-support analyst

\bibliographystyle{plainnat}
\bibliography{references}

\end{document}